# A one-dimensional discrete Boltzmann model for detonation and an abnormal detonation phenomenon

Yudong Zhang[1,2], Aiguo Xu[*,2,3], Guangcai Zhang[2], Zhihua Chen[1], Yangjun Ying[2]

*1 Key Laboratory of Transient Physics, Nanjing University of Science and Technology, Nanjing 210094, China*

*2 Laboratory of Computational Physics, Institute of Applied Physics and Computational Mathematics, Beijing, 100088, China*

*3 Center for Applied Physics and Technology, MOE Key Center for High Energy Density Physics Simulations, College of Engineering, Peking University, Beijing 100871, China*

**Abstract:** A one-dimensional discrete Boltzmann model for detonation simulation is presented. Instead of by numerical solving Navier-Stokes equations, this model obtains the information of flow field through numerical solving specially discretized Boltzmann equation. Several classical examples are simulated to validate the new model. Based on the new model, the influence of negative temperature coefficient of reaction rate on detonation is further investigated. It is found that an abnormal detonation with two wave heads periodically appears under negative temperature coefficient. The causes of the abnormal detonation are analyzed. One cycle of the periodic abnormal detonation and its development process are discussed.

## 1 Introduction

Detonation is one kind violent combustion mode accompanied with a large amount of heat release within a short time [1]. It can be treated as one kind shock wave driven by chemical reaction and propagates with a supersonic speed [2]. Detonation is closely related to the energy use and production safety. In some cases, it is necessary to generate detonation waves to improve the utilization efficiency of fuel. Because detonation possesses the isovolumetric characteristic during chemical reaction, it has a higher mechanical efficiency than the general combustion mode [3]. Based on detonation, several kinds of aero-engines have been investigated recently such as Pulse Detonation Engine [4], Rotating Detonation Engine [5], Oblique Detonation ramjet-in-Tube [6], etc. While in other cases, we should avoid the formation of detonation as far as possible or lower its destructiveness, such as in Coal Mines[7]. Detonation is so important for our life and industrial production. However, there still exist much unknown for its deep formation and propagation mechanisms [3][7].

---

[*] Corresponding author at: Laboratory of Computational Physics, Institute of Applied Physics and Computational Mathematics, Beijing 100088, China ；Xu_Aiguo@iapcm.ac.cn

It has been well known that combustion and detonation is a complex chemical reaction process with various non-equilibrium behaviors including flow non-equilibrium, thermodynamic non-equilibrium and chemical reaction non-equilibrium [9][10]. Traditional methods of investigating detonation are mainly based on using Navier-Stokes (NS) equations to describe the flow process and using phenomenological reaction rate formula to describe the reaction process. Of course, great progress has been made on the studies of detonation by the traditional method, especially in recent years [3][9]. However, NS equations themselves are not sufficient in describing the flow non-equilibrium. The coefficients of viscosity and heat conduction are generally calculated by empirical formula, such as Sutherland equation, or measured by experiment [11][12]. This method is not accurate enough when simulating the flow phenomena with strong non-equilibrium characteristics. Compared with NS equations, Boltzmann equation is more fundamental. Boltzmann equation stems from the non-equilibrium statistic mechanics, instead of fluid mechanics. It is a mesoscale model while NS equations belong to macro-scale model. By means of the Chapman-Enskog analysis, a well-known multi-scale asymptotic expansion, the Euler equations can be obtained from the Boltzmann equation when the system is exactly in its local thermodynamic equilibrium state, and the NS equations can be obtained when the system linearly, in the Knudsen number, deviates from its local thermodynamic equilibrium states. However, when the system deviates much more from its local thermodynamic equilibrium states, NS equations will not be accurate enough and fail to capture many important flow behaviors, whereas Boltzmann equation is naturally adapt to all of the above situations.

The original Boltzmann equation is too complex to be used for numerical calculations. One typical example of using Boltzmann equation in computational fluid dynamics is the Lattice Boltzmann Model (LBM) [13], and there have been several works of using LBM in combustion simulation in recent two decades. The first LBM model for combustion simulation is presented by Succi in 1997 [14]. Subsequently, Filippova [15], Yamamoto [16], Chiavazzo [17], Chen [18] and other researchers further develop the application of LBM on simulating combustion. However, all of those previous works aim only at simulating incompressible combustion and cannot satisfy the requirements of detonation simulation which shows obvious compressible behaviors. Recently, Xu's group made some attempts in simulating high speed compressible flows using LBM and developed it into a kinetic modeling method to investigate non-equilibrium characteristics [19]-[22]. To distinguish from the LBM aiming at numerical solving partial differential equations, the kinetic modeling LBM is referred to simply as Discrete Boltzmann Method/Model (DBM) in this paper. In 2013, the first DBM model for detonation was presented [24]. Then a series of extensions have been made. For example, the Multiple-relaxation-time DBM [25] and double-distribution-function DBM [26] have been developed. However, those works are all two-dimensional model and at least 16 discrete velocities are needed. So they are not economic enough from the computational side for some cases where the main behaviors can be described by one-dimensional model.

Generally speaking, during a combustion process, many species of reactants and a large number of reactions are involved. For example, the combustion of $CH_4$ in air involves 53 species and 325 reactions [27]. The reaction rate varies with the special reaction paths. A detonation process may guide the reactions into different reaction chains because of the type of fuel, shock strength, premixing homogeneity, etc. Consequently, the global reaction rates show different behaviors for different conditions and have a non-monotonic dependence on the temperature. Although most of the reaction rate has an exponential dependence on temperature and the Arrhenius model is commonly be adopted to describe the reaction rate, the phenomenon of Negative Temperature Coefficient (NTC) in reaction rate has been observed in combustion process and has drawn great attention in recent years[9][28][29]. The existence of NTC may also cause significantly different detonation behaviors. However, possible effects of NTC on detonation have not been well studied.

In 2016, we conducted a preliminary study on the effects of NTC during detonation[30]. In that work, we found that the effect of NTC during detonation is to lower the reaction rate and prolong the ignition time. In this paper, we further study the characteristics of detonation under NTC based on the one-dimensional DBM model. An abnormal detonation phenomenon, where the detonation wave has two wave heads, is observed and carefully studied.

**2 Model and verification**

2.1 Discrete Boltzmann Model and chemical reaction rate model

Below is the description of the one-dimensional DBM. The evolution of the distribution function $f_i$ for the discrete velocity $v_i$ is governed by Eq. (1):

$$\frac{\partial f_i}{\partial t} + v_i \frac{\partial f_i}{\partial x} = -\frac{1}{\tau}(f_i - f_i^{*eq}) \tag{1}$$

The subscript $i$ of $f_i$ indicates the index of discrete velocity whose speed is $v_i$. The variables, $t$ and $x$, are the time and spatial coordinates, respectively. $\tau$ is the relaxation parameter and $f_i^{*eq}$ is the local equilibrium distribution function containing the effects of chemical reaction. $f_i^{*eq}$ can be calculated by

$$f_i^{*eq}(\rho, u, T) = f_i^{eq}(\rho, u, T + \frac{1}{\gamma - 1}\tau\rho\omega Q) \tag{2}$$

where $\rho, u$ and $T$ are, respectively, the density, velocity and temperature. $\gamma$ is the ratio of special heats. $\omega$ is the chemical reaction rate, and $Q$ is the amount of heat release per unit mass of fuel. While $f_i^{eq}$ can be solved by the following seven equations:

$$\sum_i f_i^{eq} = \rho \tag{3}$$

$$\sum_i f_i^{eq} v_i = \rho u \tag{4}$$

$$\sum_i f_i^{eq}(\frac{v_i^2 + \eta_i^2}{2}) = \rho(\frac{1+n}{2}T + \frac{u^2}{2}) \tag{5}$$

$$\sum_i f_i^{eq} v_i^2 = \rho(T + u^2) \tag{6}$$

$$\sum_i f_i^{eq}(\frac{v_i^2 + \eta_i^2}{2})v_i = \rho u(\frac{1+n+2}{2}T + \frac{u^2}{2}) \tag{7}$$

$$\sum_i f_i^{eq} v_i^3 = \rho u(3T + u^2) \tag{8}$$

$$\sum_i f_i^{eq}(\frac{v_i^2 + \eta_i^2}{2})v_i^2 = \rho T(\frac{1+n+2}{2}T + \frac{u^2}{2}) + \rho u^2(\frac{1+n+4}{2}T + \frac{u^2}{2}) \tag{9}$$

where $\eta_i$ is a free parameter introduced to describe the *n* extra degrees of freedom corresponding to molecular rotation and/or vibration[20]. In this paper $\eta_i = \eta_0$ for *i*=1,…,4 and $\eta_i = 0$ for *i* =4,…,7.

From multi-scale asymptotic expansion, we know that NS equations with chemical reaction can be de deduced from Eq.(1) under the condition of Eqs.(3)-(9). Then, in order to solve the above seven equations, at least seven velocities are needed. The discrete velocities model adopted in this paper is schematically shown in Figure 1. The corresponding values of discrete velocities are listed in Table 1.

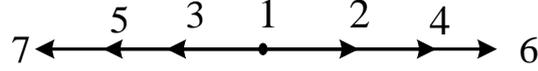

Fig. 1 Discrete velocities model for one-dimensional DBM

Table 1   Value of discrete velocities

| **Index** | $v_1$ | $v_2$ | $v_3$ | $v_4$ | $v_5$ | $v_6$ | $v_7$ |
|---|---|---|---|---|---|---|---|
| **value** | 0 | $c_0$ | $-c_0$ | $\sqrt{2}c_0$ | $-\sqrt{2}c_0$ | $2c_0$ | $-2c_0$ |

The chemical reaction rate is described by Eq.(10):

$$\omega = \begin{cases} k\lambda(1-\lambda), & T \geq T_{th} \\ 0 & T < T_{th} \end{cases} \quad (10)$$

where k indicates reaction rate constant and $\lambda$ is reaction process. $\lambda = 0$ indicates that reaction has not started and $\lambda = 1$ indicates that reaction is completed. $T_{th}$ is defined as ignition temperature and there is no reaction when the temperature of reaction region is lower than this threshold value.

2.2 model verification

In order to validate the new DBM model, several typical benchmarks are carried out. Firstly, two shock wave problems without chemical reaction are simulated and compared with the Riemann solutions including Sod shock tube and Colella's explosion wave problem. Then one-dimensional self-sustainable stable detonation is simulated and compared with CJ solutions.

(1)Sod shock tube test

The initial conditions are

$$\begin{cases} (\rho, u, T)_L = (1, 0, 1), \\ (\rho, u, T)_R = (0.125, 0, 0.8). \end{cases} \quad (11)$$

The comparisons of the DBM results and Riemann exact solution at $t = 0.22$ are shown in Figure 2. Simulation is carried under the condition that $\tau = 2 \times 10^{-5}$, $\Delta x = 2 \times 10^{-4}$, $Nx = 5000$, $\Delta t = 5 \times 10^{-6}$, $c_0 = 1.2, \eta_0 = 3$. Besides, $n = 4$ so that $\gamma = \dfrac{n+1+2}{n+1} = 1.4$. From the Fig.2, we can see that the DBM results are well consistent with the Riemann solutions.

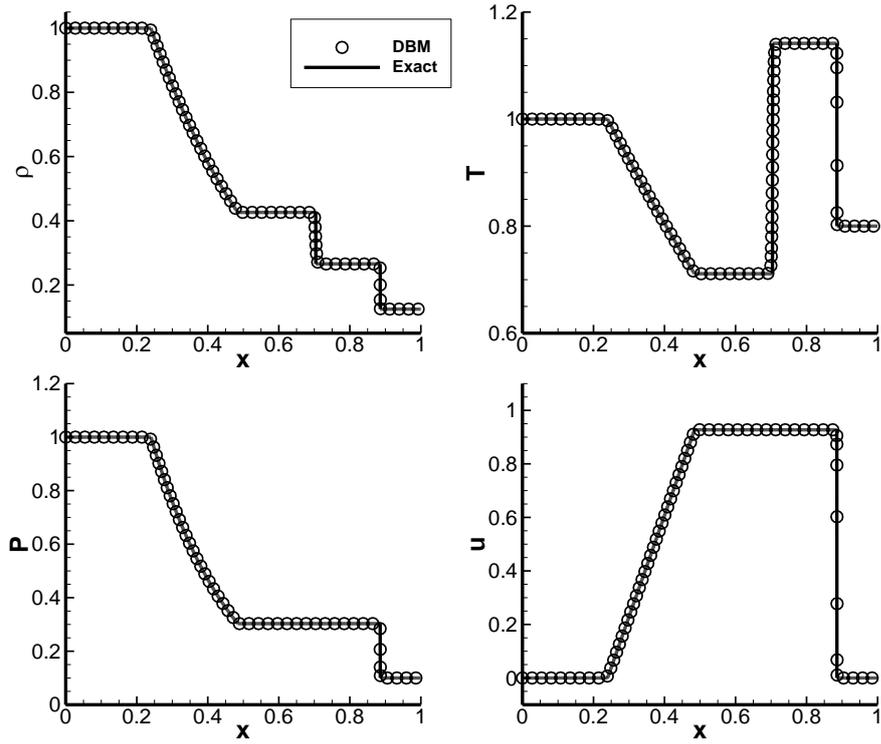

Fig. 2 Comparisons between DBM results and exact solutions for Sod shock tube test.

(2)Colella's explosion wave test
The initial conditions are

$$\begin{cases} (\rho,u,T)_L = (1,0,1000), \\ (\rho,u,T)_R = (1.0,0,0.01). \end{cases} \quad (12)$$

Simulation is carried under the condition that $\tau = 1\times 10^{-5}$, $\Delta x = 2\times 10^{-3}$, $Nx = 2000$, $\Delta t = 5\times 10^{-6}$, $c_0 = 20$, $\eta_0 = 16$. Besides, $n = 4$ so that $\gamma = \dfrac{n+1+2}{n+1} = 1.4$. The simulation results at $t = 0.025$ are shown in Figure 3 from which we can see that the new DBM model is applicable to flows with very high ratios of temperature (up to $10^5$).

(3)Self-sustainable stable detonation
As the last test, one-dimensional self-sustainable stable detonation is simulated. Considering that there is a rigid tube full of premixed combustible gas. The Initial condition in this tube is set as follows:

$$\begin{cases} (\rho,u,T,\lambda)_L = (1.38837, 0.57735, 1.57856, 1), \\ (\rho,u,T,\lambda)_R = (1,0,1,0). \end{cases} \quad (13)$$

Parameters are set as $\tau = 2\times 10^{-5}$, $\Delta x = 2\times 10^{-4}$, $Nx = 5000$, $\Delta t = 5\times 10^{-6}$, $c_0 = 2$, $\eta_0 = 2$. Besides, $n = 4$ so that. $\gamma = 1.4$. The left boundary is set as fixed wall while the right set as outflow.

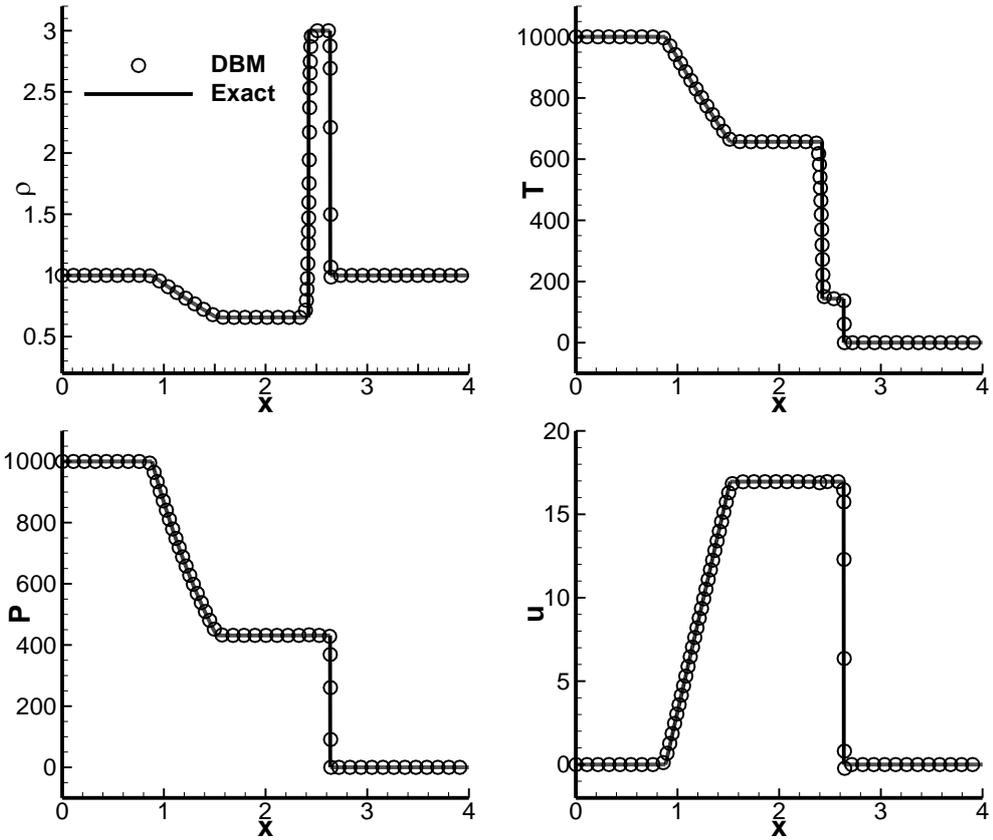

Fig. 3 Comparisons between DBM results and exact solutions for Colella's explosion wave test.

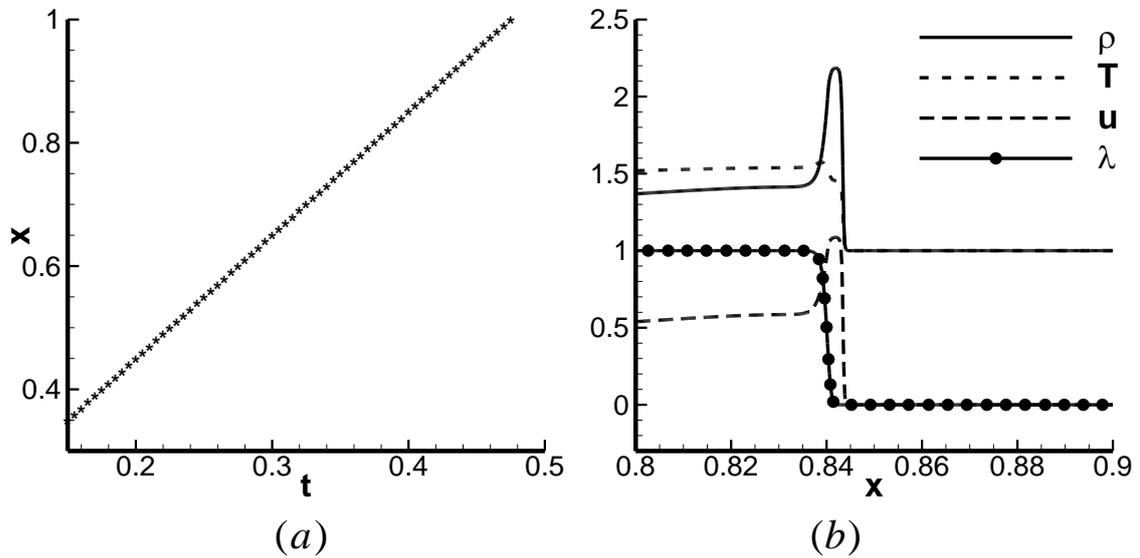

Fig. 4 Simulation results of self-sustain stable detonation.(a) the position of von-Neumann peak of pressure. (b) the spatial distribution of macro quantities at $t$=0.4.

The results are shown in Figure 4. We can see that the self-sustain detonation with a stable wave speed is formed. Then we further compare the macro quantities behind the detonation wave at sound velocity surface with the CJ theoretical value. The corresponding values are shown in Table 2. The relative errors are all less than 0.055.

So we can conclude that the new DBM can be used to investigate the one-dimensional detonation problem.

Table 2 Comparisons between DBM results and CJ theoretical value for stable detonation

|  | DBM results | CJ theoretical value | Relative errors |
|---|---|---|---|
| **D(wave speed)** | 2.00514 | 2.06395 | 0.0285 |
| $\rho$ | 1.37528 | 1.38837 | 0.0131 |
| u | 0.54598 | 0.57735 | 0.0543 |
| T | 1.52146 | 1.57856 | 0.0362 |

## 3 Results and discussion:

### 3.1 Simulation Results

In order to investigate the NTC of chemical reaction rate, we adopt the following formula to describe the dependence of $k$ (in Eq.10) on temperature which was presented in Ref.[30].

$$k(T) = a + b\left(\frac{1}{3}T^3 - \frac{1}{2}(T_1+T_2)T^2 + T_1T_2T\right) \tag{14}$$

with

$$a = \frac{-h_2T_1^3 + 3h_2T_1^2T_2 - 3h_1T_1T_2^2 + h_1T_2^3}{(T_1-T_2)^2} \tag{15}$$

$$b = -\frac{6(h_1-h_2)}{(T_1-T_2)^3} \tag{16}$$

where $h_1$ and $h_2$ are the peak and valley value of k, respectively. $T_1$ and $T_2$ are temperature corresponding to $h_1$ and $h_2$. In this work, we set $h_1 = 2000$, $h_2 = 10$, $T_1 = 1.14$, $T_2 = 1.45$. Besides, the ignition temperature in Eq. (10) is set as $T_{th} = 1.1$. The curve of $k$ with temperature $T$ is shown in Figure 5.

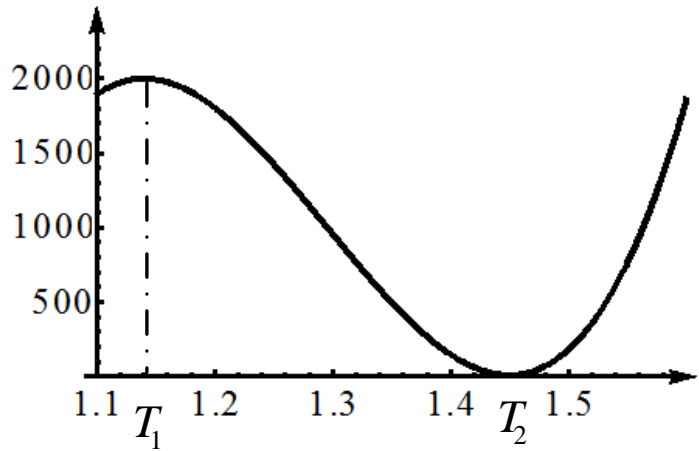

Fig. 5 Curve of k with temperature (T).

Under this chemical reaction case, an abnormal detonation phenomenon is obtained, which is shown in Figure 6. Fig. 6(a) shows a normal stable detonation wave with a constant speed and a stable waveform, while Fig. 6(b) shows the abnormal detonation under the chemical reaction rate as shown in Figure 5. For this abnormal detonation, there is no constant wave speed and the waveform changes periodically. From Fig.6(b) we can see the evolution of the waveform in one cycle. Firstly, at a certain time, a local hotspot generates behind the wave head. Then a local detonation wave appears and develops with a faster speed than its downstream wave. So the new formed detonation wave chases the front wave and finally the two waves are combined into an over-driven detonation wave. However, over-driven detonation wave cannot self-sustain, so the wave velocity is gradually reduced until it reaches to the CJ wave velocity. After that, a local hotspot generates again and a new cycle begins.

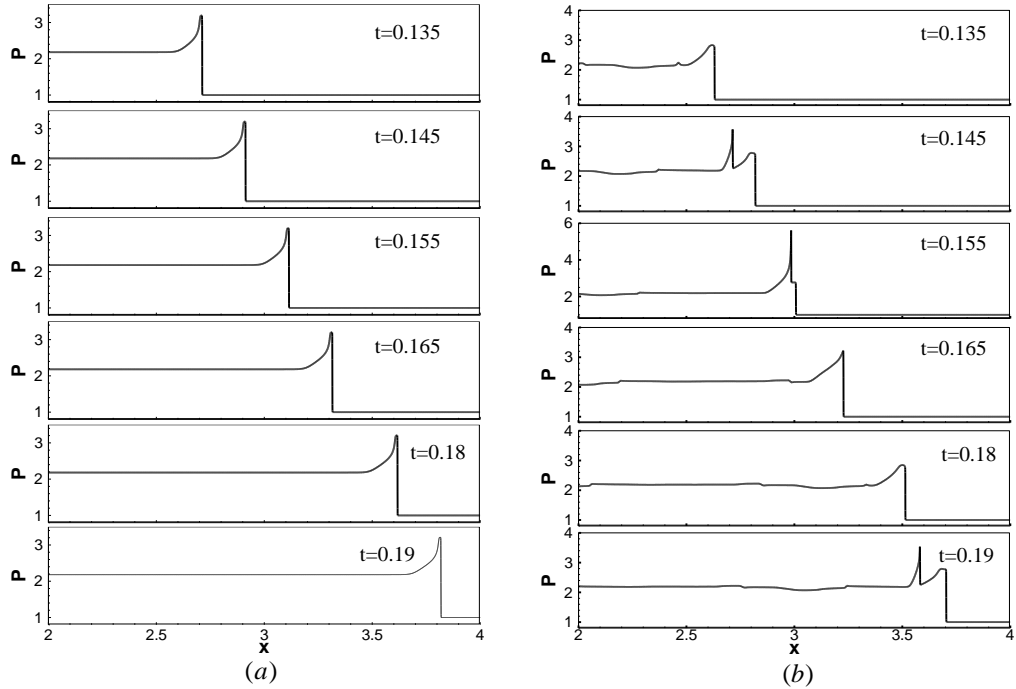

Fig. 6 Comparisons between (a) the normal detonation wave and (b) the abnormal detonation wave.

**3.2 Discussion**

In this section, we will discuss the causes of formation of the abnormal detonation wave. For convenience, we roughly divide the reaction into three stages according to the temperature. Those three stages are denoted as S1, S2 and S3, respectively, as shown in Figure 7. The first stage (S1) has a fast reaction rate but at low temperature because of NTC. The second stage (S2) has a slow reaction rate at specific temperature range. The third stage (S3) has a fast reaction rate at high temperature and increases quickly with the increasing temperature.

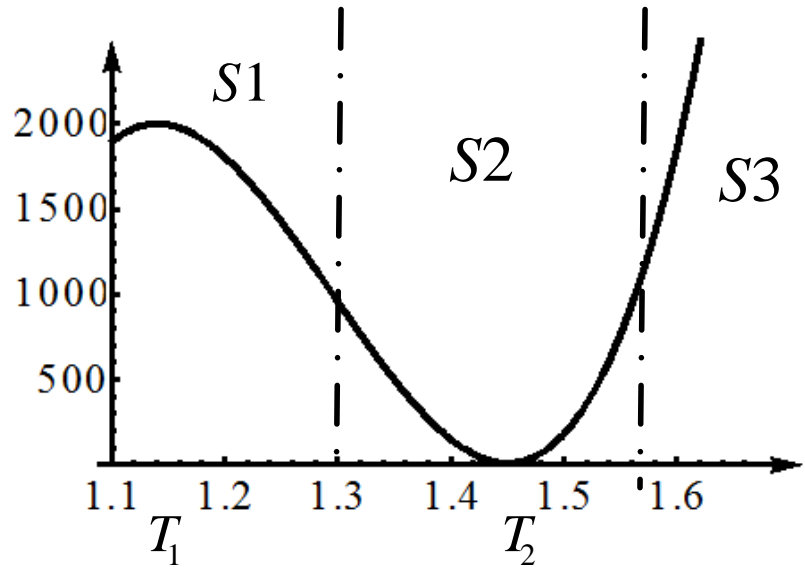

Fig. 7 Three stages of chemical reaction rate.

The development of abnormal detonation is shown in Figure 8. At $t=0.12$ (Fig. 8a), the detonation wave is still a normal detonation followed by a long sparse wave region. The temperature behind the wave head has not reach the third stages (S3) in Figure 7. So the chemical reaction has only the first two stages, S1 and S2. Then at $t=0.13$ (Fig. 8b), a local hotspot appears in the sparse wave region and the temperature in this hotspot achieves to S3, so the third stage of chemical reaction appears. Because it has a much faster reaction rate in S3, a localized detonation wave is formed and developed quickly. The localized detonation moves forward and obtains more fuel. So, it continues to be strengthened and has an increasing wave speed. From Fig.8(c) we can see S3 gradually widens while S2 becomes gradually narrower. At $t=0.145$, the new formed detonation wave almost catches up with the front detonation wave. At this time, S2 is nearly disappeared (Fig. 8d). After that, two waves are merged and the overdriven detonation wave occurs.

Figure 9(a) shows the overdriven detonation wave at $t=0.155$, which has a wave speed faster than the CJ detonation wave speed. At this time, almost all of the chemical reactions are in S3. However, overdriven detonation cannot self-sustain so sparse waves would be gradually formed behind the wave head. At $t=0.165$ (Fig. 9b), the temperature behind the detonation wave front begins to drop due to the effect of the sparse wave. When the temperature behind the wave head declines to S2, the second stage chemical reaction occurs. Then the sparse waves behind the wave head continue developing and temperature behind the wave keeps dropping. When the temperature declines to S1, the first stage reaction occurs. As more and more fuel react in S1 and S2, reaction occur in S3 gradually decrease (Fig. 9c and Fig. 9d). With the decrease of chemical reaction rate, the detonation wave speed gradually decreases to CJ detonation wave speed. At the next time, a local hotspot appears and a new local detonation wave is developed again. So, the above process is repeated. The whole process can be summarized by Figure 10.

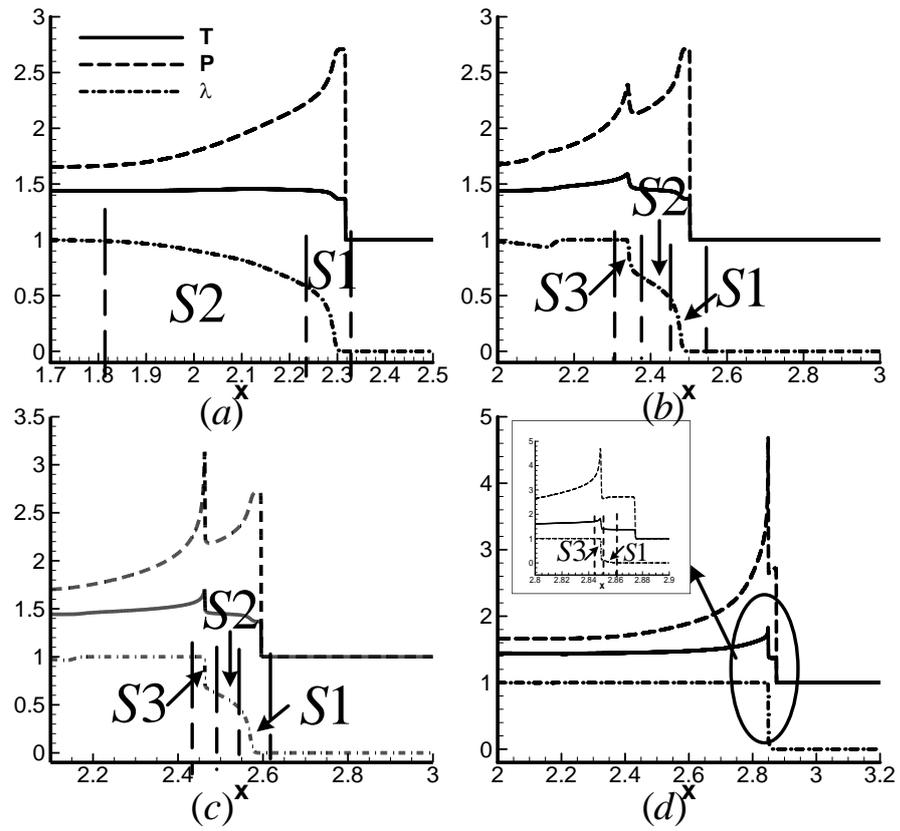

Fig. 8 Detonation waveforms at (a) $t$=0.12 (b) $t$=0.13 (c) $t$=0.135 (d) $t$=0.15.

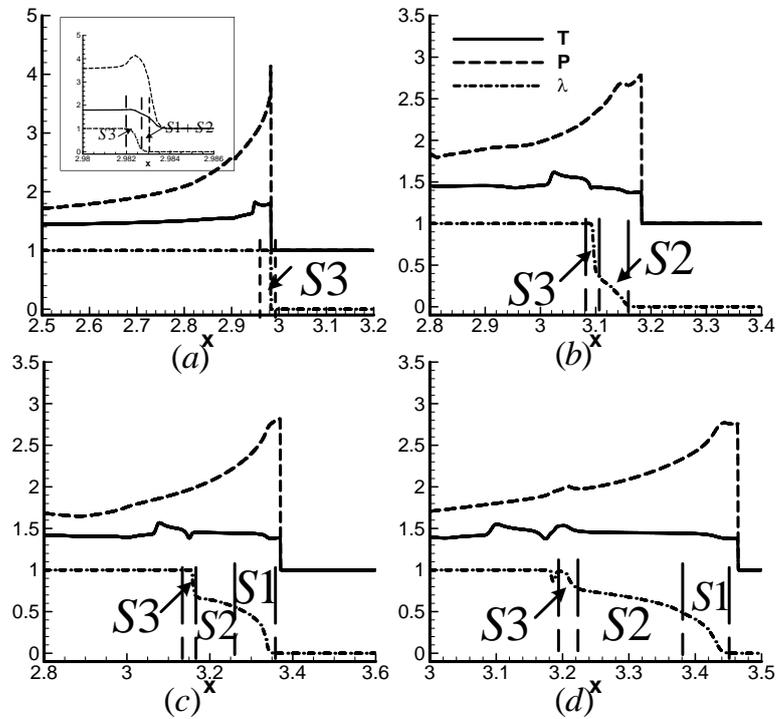

Fig. 9 Detonation waveforms at (a) $t$=0.155 (b) $t$=0.165 (c) $t$=0.175 (d) $t$=0.18.

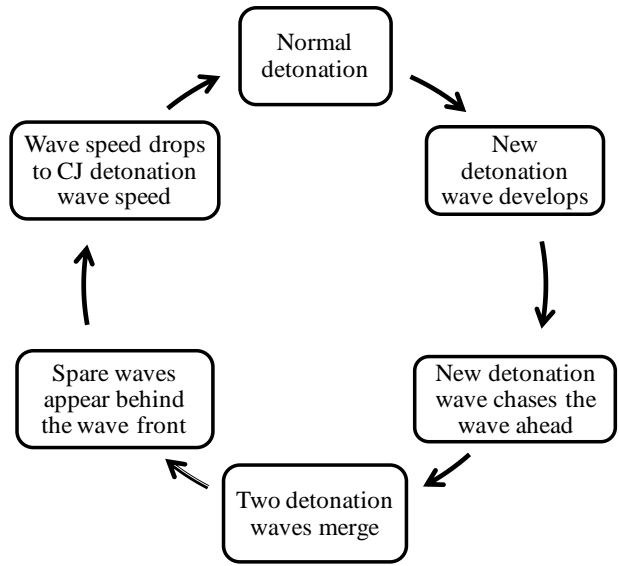

Fig. 10   Schematic diagram of abnormal detonation development process

**4 Conclusions:**

In this work, we present a one-dimensional DBM for detonation. The validity of the new model is verified by three tests. Based on the new model, an abnormal detonation phenomenon is presented and its development process is analyzed. It is found that the main reason for the abnormal detonation is that the chemical reaction has three stages, namely S1, S2 and S3. For a normal detonation, the fuel mainly reacts in S1 and S2. At a certain time, a local hotspot is formed and a new detonation appears behind the old detonation wave front. Because the new detonation wave has a faster speed that the wave ahead, it catches up with the front wave and two waves merge at last. Then the speed of the detonation wave begins to slow down until it reaches CJ detonation wave speed. After that, the local hotspot appears again and the previous process reappears.

**References**

[1]   C. L. Mader. Numerical modeling of detonations [M]. University of California Press, 1979.

[2]   J. S. Sun,   J. S. Zhu, Theory of detonation physics [M], National Defence Industry Press, Beijing, 1995

[3]   Z. Jiang, H. Teng, Y. Liu. Some research progress on gaseous detonation physics [J]. Advances in Mechanics, 2012, 42(2): 183-183.

[4]   K. Kailasanath. Recent Developments in the Research on Pulse Detonation Engines [J]. Aiaa Journal, 2003, 41(2):145-159.

[5]   Y. T. Shao, M. Liu, J. P. Wang. Numerical investigation of rotating detonation engine propulsive performance [J]. Combustion Science and Technology, 2010, 182(11-12): 1586-1597.

[6]   D. W. Bogdanoff, D. C. Brackett. Computational investigation of oblique detonation ramjet-in-tube concepts [J]. Journal of Propulsion and Power, 1989, 5(3): 276-281.

[7]   H. Yu, S. Zhang. Discussion on Occurrence Regularity and Countermeasures of Gas Accidents in Our Coal Mines in Recent Years [J]. Mining Safety & Environmental Protection, 2016, 43(1):108-110.

[8]   J. E. Shepherd. Detonation in gases [J]. Proceedings of the Combustion Institute, 2009, 32(1): 83-98.

[9]   Y. G. Ju. Recent progress and challenges in fundamental combustion research [J]. Advances in Mechanics, 2014, 44(20): 201402.

[10]   A. Xu, G. Zhang G, Y. Ying. Progress of discrete Boltzmann modeling and simulation of combustion system

[J]. Acta Phys. Sin, 2015, 64(18): 184701.
[11]  D. Fu, Y. Wang, Computational aerodynamics [M]. Aerospace press, 1994
[12]  J. D. Anderson, J. Wendt. Computational fluid dynamics [M]. New York: McGraw-Hill, 1995.
[13]  S. Succi. The lattice Boltzmann equation: for fluid dynamics and beyond [M]. Oxford university press, 2001.
[14]  S. Succi, G. Bella, F. Papetti. Lattice kinetic theory for numerical combustion [J]. Journal of scientific computing, 1997, 12(4): 395-408.
[15]  O. Filippova, D. Haenel. A novel numerical scheme for reactive flows at low Mach numbers [J]. Computer physics communications, 2000, 129(1): 267-274.
[16]  K. Yamamoto, N. Takada, M. Misawa. Combustion simulation with Lattice Boltzmann method in a three-dimensional porous structure [J]. Proceedings of the Combustion Institute, 2005, 30(1): 1509-1515.
[17]  E. Chiavazzo, I. V. Karlin, A. N. Gorban, et al. Coupling of the model reduction technique with the lattice Boltzmann method for combustion simulations [J]. Combustion and Flame, 2010, 157(10): 1833-1849.
[18]  S. Chen, J. Mi, H. Liu, et al. First and second thermodynamic-law analyses of hydrogen-air counter-flow diffusion combustion in various combustion modes[J]. international journal of hydrogen energy, 2012, 37(6): 5234-5245.
[19]  A. Xu, G. Zhang, Y. Gan, et al. Lattice Boltzmann modeling and simulation of compressible flows [J]. Frontiers of Physics, 2012, 7(5): 582-600.
[20]  Y. Gan, A. Xu, G. Zhang, et al. Lattice BGK kinetic model for high-speed compressible flows: Hydrodynamic and non-equilibrium behaviors [J]. EPL (Europhysics Letters), 2013, 103(2): 24003.
[21]  C. Lin, A. Xu, G. Zhang, et al. Polar-coordinate lattice Boltzmann modeling of compressible flows[J]. Physical Review E, 2014, 89(1): 013307.
[22]  A. Xu, G. Zhang, Y. Li, et al. Modeling and Simulation of Non-equilibrium and Multiphase Complex Systems: Lattice Boltzmann kinetic Theory and Application [J]. Progress in Physics, 2014, 34(3): 136-167.
[23]  A. Xu, G. Zhang, and Y. Gan. Progress in studies on discrete Boltzmann modeling of phase separation process [J], Mechanics in Engineering., **38** (4), 361-374 (2016).
[24]  B. Yan, A. Xu, G. Zhang, et al. Lattice Boltzmann model for combustion and detonation [J]. Frontiers of Physics, 2013, 8(1): 94-110.
[25]  A. Xu, C. Lin, G. Zhang, et al. Multiple-relaxation-time lattice Boltzmann kinetic model for combustion[J]. Physical Review E, 2015, 91(4): 043306. 38(4)
[26]  C. Lin, A. Xu, G. Zhang, et al. Double-distribution-function discrete Boltzmann model for combustion[J]. Combustion and Flame, 2016, 164: 137-151.
[27]  M. Faghih, X. Gou, Z. Chen. The explosion characteristics of methane, hydrogen and their mixtures: A computational study[J]. Journal of Loss Prevention in the Process Industries, 2015, 40(22):131-138.
[28]  G. I. Ksandopulo. Staging, negative temperature coefficient of the reaction rate and bifurcation in the monofront of hydrocarbon flames[J]. Russian Journal of Physical Chemistry B, 2011, 5(4):701-711.
[29]  J. Pan, H. Wei, G. Shu, et al. The role of low temperature chemistry in combustion mode development under elevated pressures[J]. Combustion & Flame, 2016, 174:179–193.
[30]  Y. Zhang, A. Xu, G. Zhang, et al. Kinetic modeling of detonation and effects of negative temperature coefficient[J]. Combustion & Flame, 2016, 173:483-492.